# Coalescence of spectral singularities and phase diagrams for one-dimensional *PT* symmetric photonic crystals

Kun Ding, Z. Q. Zhang, and C. T. Chan[†]

*Department of Physics and Institute for Advanced Study, The Hong Kong University of Science and Technology, Clear Water Bay, Hong Kong*

† Corresponding E-mail: phchan@ust.hk

**Abstract**

Non-Hermitian systems with parity-time (*PT*) symmetric complex potentials can exhibit a phase transition when the degree of non-Hermiticity is increased. Two eigenstates coalesce at a transition point, which is known as the exceptional point (EP) for a discrete spectrum and spectral singularity for a continuous spectrum. The existence of an EP is known to give rise to a great variety of novel behaviors in various fields of physics. In this work, we study the complex band structures of one-dimensional photonic crystals with *PT* symmetric complex potentials by setting up a Hamiltonian using the Bloch states of the photonic crystal without loss or gain as a basis. As a function of the degree of non-Hermiticity, two types of *PT* symmetry transitions are found. One is that a *PT*-broken phase can re-enter into a *PT*-exact phase at a higher degree of non-Hermiticity. The other is that two spectral singularities, one originating from the Brillouin zone center and the other from the Brillouin zone boundary, can coalesce at some *k*-point in the interior of the Brillouin zone and create a singularity of higher order. Furthermore, we can induce a band inversion by tuning the filling ratio of the photonic crystal, and we find that the geometric phases of the bands before and after the inversion are independent of the amount of non-Hermiticity as long as the *PT*-exact phase is not broken. The standard concept of topological transition can hence be extended to non-Hermitian systems.

**Section I. Introduction**

Non-Hermitian systems can exhibit many interesting phenomena [1,2]. The extension of Hermitian Hamiltonians to non-Hermitian ones can be achieved mathematically by various ways, such as by introducing a complex potential or adding asymmetric losses to the system [2,3]. It is also recognized that if the complex potential has parity-time (*PT*) symmetry, the system will possess real eigenvalues at small non-Hermiticity and when the non-Hermiticity is sufficiently large, the system will experience a phase transition at an exceptional point (EP).

In optics, the dielectric constant often plays the role of potential. If the material possesses the property of $\varepsilon(x) = \varepsilon^*(-x)$, the system is *PT*-symmetric because the system restore itself after simultaneous parity and time-reversal operations. Such systems have been extensively studied recently and a great variety of interesting phenomena have been discovered, including unusual beam dynamics within paraxial approximations [4-6], lasing effect [7-9], unidirectional transmission [10-12], negative refraction [13], single particle sensors [14] and others [15-20]. Since it is not easy to achieve *PT*-symmetry experimentally, several ways have been proposed to avoid the use of gain in a system. For example, a system with asymmetric loss is equivalent to a *PT*-symmetric system with a background of uniform loss, and so no gain is required to achieve an EP [3, 21-24]. In a system with many coupled modes, the emergence of multiple EPs and their interactions can occur under system parameter variation [25-27]. Their interactions and the associated topological properties have been studied in both microwave cavities [25,26] and acoustics systems [27]. For systems with continuous spectra, various *PT*-symmetric photonic crystals (PCs) have been considered by studying complex band structures. In such periodic systems, an EP normally occurs when a gap is closed at some high symmetry point of the Brillouin zone. Thus, this kind of EP is also called spectral singularities [28-33]. For example, in the *PT*-symmetric plasmonic waveguides, EPs can close the gap formed by two branches of surface plasmon polariton [28]. EPs can also close band gaps in *PT*-symmetry honeycomb PCs [29,30]. A ring of EPs can exist near a Dirac-like cone in 2D PCs [33].

In this paper, we investigate the evolution of complex band structures in a 1D *PT*-symmetric PC as the amount of non-Hermiticity increases continuously. We found two types of *PT* phase diagram. In the first type, a *PT*-broken phase can re-enter into the *PT*-exact phase when the non-Hermiticity is further increased. In the other type, two EPs (or spectral singularities) coalesce and produce a singularity of higher order. In addition, we tune the filling ratio of the photonic crystal to induce a band inversion. We show that the Zak phases of the bands before and after the inversion are independent of the amount of non-Hermiticity before the *PT*-exact phase is broken. This generalizes the concept of topological transition from Hermitian systems to non-Hermitian systems. The paper is organized as follows. In Section II, we formulate the model Hamiltonian to study the complex band structures of 1D *PT*-symmetric PCs. The numerical results of complex band structures together with the two types of phase diagram are presented in Section III. The calculation of Zak phase for non-Hermitian systems are presented in Section IV for both separated and entangled bands. Conclusions are drawn in Section V.

**Section II. Formulation of Model Hamiltonian**

The system studied is a 1D PC with a *PT*-symmetric complex dielectric constant. Each unit cell of the PC consists of four dielectric layers as shown in Fig. 1(a). The dielectric constants of these layers are $\varepsilon_b - i\Gamma_b$, $\varepsilon_a + i\Gamma_a$, $\varepsilon_a - i\Gamma_a$, and $\varepsilon_b + i\Gamma_b$, respectively. In the absence of gain or loss, this system is a simple superlattice formed by alternating layers of A and B with dielectric constants $\varepsilon_a$ and $\varepsilon_b$, respectively. By taking the middle plane of the A layer as the origin of the system, we see that $\varepsilon(x) = \varepsilon^*(-x)$ and the system is *PT*-symmetric, i.e., the system restores itself after simultaneous parity and time reversal operations. In this work, we set $\Gamma_a = \Gamma_b = \Gamma$ for simplicity.

The electric field in the layer satisfies the following Helmholtz equation [34],

$$\left[\frac{d^2}{dx^2}+\left(\frac{\omega}{c}\right)^2\varepsilon(x)\right]E(x)=0. \qquad (2.1)$$

The eigenfrequency and eigenfunction of a Bloch state can be obtained using the transfer matrix method (TMM), which relates the fields between two adjacent unit cells [35], i.e.,

$$\mathbf{M}\begin{pmatrix}a_l\\b_l\end{pmatrix}=e^{ik\Lambda}\begin{pmatrix}a_l\\b_l\end{pmatrix}, \qquad (2.2)$$

where $\mathbf{M}=\prod_{p=1}^{4}\mathbf{M}_p$ is total transfer matrix for one unit cell, $a_l, b_l$ are the field amplitudes of forward-going and backward-going waves in the $l$-th layer, $k$ is Bloch wave vector and $\Lambda$ is the lattice constant. In the absence of loss or gain in the system ($\Gamma=0$), the dispersion relations obtained from Eq. (2.2) are shown by solid lines in Fig. 1(b), which includes 6 lowest bands and 5 lowest gaps. In the calculation, we have chosen $\varepsilon_a=3.8$, $\varepsilon_b=1$ and $\mu_a=\mu_b=1$. The thicknesses of the A and B layers are $d_a=0.42\Lambda$, and $d_b=0.58\Lambda$, respectively. To calculate the complex band structures when $\Gamma\neq 0$, we construct a model Hamiltonian using the Bloch states obtained at any fixed value of $k$ with $\Gamma=0$ as a basis for the case when $\Gamma\neq 0$. The Bloch states for $\Gamma=0$ can be expressed as $E_{nk}^{(0)}(x)=u_{nk}(x)e^{ikx}$, where $n$ denotes the band index and is a positive integer. The periodic function $u_{nk}(x)$ and the corresponding eigenfrequency $\omega_n(k)$ can be obtained from Eq. (2.2). Since the system possesses inversion symmetry when $\Gamma=0$, $u_{nk}(x)$ is either symmetric or anti-symmetric at the Brillouin center or boundaries. In Fig. 1(b), we also show the symmetry of the Bloch states at band edges with the solid red circles representing even modes, i.e., $u_{nk}(-x)=u_{nk}(x)$ and the solid grey circles representing odd modes, i.e., $u_{nk}(-x)=-u_{nk}(x)$, where $k\Lambda/2\pi=0$ or $\pm 0.5$. To construct the model Hamiltonian for any given $k$, we express the Bloch wavefunctions of the *PT*-symmetric system as $E_{nk}^{PT}(x) = \tilde{u}_{nk}(x)e^{ikx}$ with $\tilde{u}_{nk}(x) = \sum_{m=1}^{+\infty} \alpha_{n,mk} u_{mk}(x)$.

Substituting this expansion into Eq.(2.1), we arrive

$$\sum_{n=1}^{+\infty} \alpha_{m,nk} \left\{ \left[ \left(\frac{\tilde{\omega}_m}{c}\right)^2 - \left(\frac{\omega_n(k)}{c}\right)^2 \right] \varepsilon_r(x) + \left(\frac{\tilde{\omega}_m}{c}\right)^2 i\varepsilon_i(x) \right\} u_{nk}(x) = 0, \qquad (2.3)$$

where $\varepsilon_r$ and $\varepsilon_i$ are, respectively, the real and imaginary parts of the dielectric constant, and $\omega_n(k)$ and $\tilde{\omega}_m$ are, respectively, the eigenfrequencies when $\varepsilon_i(x)=0$ and $\varepsilon_i(x)\neq 0$. Multiplying Eq. (2.3) by $u^*_{n'k}(x)$ and then integrating within a unit cell, we obtain

$$\sum_{n=1}^{+\infty} \left(\frac{\omega_n(k)}{c}\right)^2 \left(\bar{\varepsilon}_r(k)\right)_{n'n} \alpha_{m,nk} = \left(\frac{\tilde{\omega}_m}{c}\right)^2 \sum_{n=1}^{+\infty} \left[ \left(\bar{\varepsilon}_r(k)\right)_{n'n} + i\left(\bar{\varepsilon}_i(k)\right)_{n'n} \right] \alpha_{m,nk}, \qquad (2.4)$$

in which $\left(\bar{\varepsilon}_{r,i}(k)\right)_{n'n} = \int dx\, u^*_{n'k}(x)\varepsilon_{r,i}(x)u_{nk}(x)$. Eq. (2.4) can be rewritten as an eigenvalue problem

$$H \cdot \tilde{p} = H_1^{-1} \cdot H_2 \cdot \tilde{p} = \left(\frac{\omega}{c}\right)^2 \tilde{p}, \qquad (2.5)$$

where $\tilde{p} = (\alpha_{1k},\cdots,\alpha_{nk},\cdots)^T$, $(H_1)_{n'n} = \left(\bar{\varepsilon}_r(k)\right)_{n'n} + i\left(\bar{\varepsilon}_i(k)\right)_{n'n}$, and $(H_2)_{n'n} = \left(\frac{\omega_n(k)}{c}\right)^2 \left(\bar{\varepsilon}_r(k)\right)_{n'n}$. Here $H$ can be considered as an equivalent non-Hermitian Hamiltonian for the *PT*-symmetric PCs and the eigenfrequency $\omega$ is complex in general. The formation of Hamiltonian (2.5) offers us an easy way to investigate the complex band structures of *PT*-symmetric PCs as a function of non-Hermiticity. It also helps us to understand the roles played by various non-Hermitian interactions $\left(\bar{\varepsilon}_i\right)_{n'n}$ in determining the complex phase diagrams. Since the gap is the smallest at Brillouin center or boundaries, a spectral singularity will first appear at these high symmetry points due to the presence of the non-Hermitian term $\left(\bar{\varepsilon}_i(k)\right)_{n'n}$ at $k\Lambda/2\pi = 0$ or $\pm 0.5$. Since $\varepsilon_i(x)$ is an odd function of $x$, $\left(\bar{\varepsilon}_i\right)_{n'n}$ is nonzero only when the modes $n'$ and $n$ have the opposite symmetries. The existence of a non-zero $\left(\bar{\varepsilon}_i(k)\right)_{n'n}$ represents an attractive interaction between $n'$ and $n$. For two adjacent modes, such attractive interaction can give rise to an EP.

**Section III. Complex Band Structures and *PT* Phase Diagrams**

As we turn on the gain and loss, some gap at the Brillouin center or boundaries will close at a certain critical value of $\Gamma$, giving rise to a spectral singularity. As the value of $\Gamma$ is further increased, the spectral singularity moves away from the Brillouin center or boundaries giving rise to a region of *PT*-broken phase in *k* space. In the meantime, some other gap will close and produces another spectral singularity. This process is shown in Fig. 2 where the complex band structures for four different values of $\Gamma$ are shown, in which the solid lines are obtained by using Eq. (2.5) including 14 lowest bands in the calculation. For comparison, we also calculate the complex band structures using a Finite Element Method (FEM) [36] as shown by open dots in Fig. 2. Excellent agreements are found for both real and imaginary parts of the eigenfrequencies for the 5 lowest bands we are concerned in this study. This verifies that the truncation of 14 bands in the model Hamiltonian is accurate enough to produce the essential physics.

At $\Gamma = 0.1$, the real and imaginary parts of the eigenfrequencies are shown, respectively, by black and blue lines in Figs. 2(a) and 2(b). We see that at this small value of $\Gamma$ all four lowest gaps remain open and the eigenfrequencies in all four bands remain real. When the value of $\Gamma$ is increased to some critical value, a spectral singularity first appears at the Brillouin center or boundaries where the gap is the smallest. With further increase of $\Gamma$, the spectral singularity moves away from the center or boundaries and form a region of *PT*-broken phase. Figs. 2(c) and 2(d) show the band structures at $\Gamma = 0.5$. It shows that the second, third and fourth gaps have already passed their respective critical $\Gamma$ for gap closing and have entered the regime of *PT*-broken phases, which are marked in the figure by green, gray and yellow solid circles, respectively. These spectral singularities are marked by the letters *M, S,* and *N* in Fig. 2(c). Since the complex band structures of our system possess mirror symmetry in *k* space, i.e. $\omega(k) = \omega^*(-k)$, the spectral singularities always appear in pairs [37].

If we further increase $\Gamma$ to 1.0, the regions of *PT*-broken phase for the second and third gaps are expanded as shown in Figs. 2(e) and 2(f). On the contrary, we note that the *PT*-broken phase created by the fourth gap disappears and the gap re-opens. The reopening of the gap is due to the attractive interaction between the fourth band and the sixth band, which is much stronger than the interaction between the fourth and fifth bands. To illustrate this, we plot in Fig. 3 the complex frequency trajectories as functions of $\Gamma$ (dotted lines) for the three band edge states of the fourth, fifth and sixth bands at $k\Lambda/2\pi = 0$. From the symmetries of these states shown in Fig. 1(b), we see that the lowest two states mix and merge first at some critical $\Gamma$ due to opposite symmetries as we stated in section II. As $\Gamma$ is further increased, the re-emergence of a *PT*-exact phase at a higher $\Gamma$ suggests a de-mixing mechanism due to the dominant attractive interaction between the lowest frequency state and the highest state, which tends to pull the lowest state out of the broken phase, restoring the *PT* symmetry of the wavefunction. Such a coalescence and re-bifurcation cannot occur in a 2x2 system. In order to verify that this mechanism is due to multiple state coupling, we construct a reduced 3x3 Hamiltonian including only these three states in Eq. (2.5). The results are shown by solid lines in Fig. 3. The qualitative agreement between the dotted and solid lines confirms that the re-entry behavior comes from interaction with the third state. Further increasing $\Gamma$ to 2.0, as shown in Figs. 2(g) and 2(h), we discover that the spectral singularities labeled as *M* and *S* coalesce with each other at some special *k*-point in the Brillouin zone denoted by *R* in Fig. 2(g) at which the imaginary parts of the eigenfrequencies bifurcate in both directions of *k* as shown in Fig. 2(h). Like the re-entry behavior, such a coalescence of spectral singularities is also a result of many-state interaction [27].

It should be pointed out that the model Hamiltonian Eq. (2.5) has two independent parameters $\Gamma$ and *k*. The coalescence of two states can be induced by varying either one. The spectral singularity hence resides in a two dimensional space $(k,\Gamma)$. In Fig. 4, we plot the trajectories of the spectral singularities *M*, *S*, *R* in (a) and *N* in (b) in the $(k,\Gamma)$ space. In Fig. 4(a), the spectral singularities *M* and *S* first appear at $k\Lambda/2\pi = 0$ and $k\Lambda/2\pi = 0.5$ when $\Gamma$ reaches its respective critical value at 0.346 and 0.371. They move towards each other when $\Gamma$ is increased, and coalesce with each other at $\Gamma = 1.82$ and wavevector $k\Lambda/2\pi = 0.31$, creating a higher order singularity labeled as *R* in Figs. 2(g) and 4(a). After the coalescence, the singularity *R* moves continuously towards the Brillouin boundaries, as shown by yellow lines in Fig. 4(a). The trajectories of *M* and *S*, as shown by blue and red lines in Fig. 4(a), respectively, divide the parameter space $(k,\Gamma)$ into a *PT*-exact phase (white region) and a *PT*-broken phase (gray region). Figure 4(b) shows a different *PT* phase diagram for the spectral singularity *N*. It firstly appears at the Brillouin center, and then move towards the Brillouin boundaries. However, the trajectories turn around at some $\pm k_{max}$ and move back to the center, forming a closed loop enclosing an island of *PT*-broken phase inside the loop and a *PT*-exact phase outside. Thus, the loop represents a ring of spectral singularities in the $(k,\Gamma)$ plane. These two types of *PT* phase diagram have distinct topological characteristics as will be discussed below.

To identify the order of singularity for points *S, M, R* and *N*, we study the phase rigidity $r_j$ for bands $j = 2, 3$ and 4 defined as $r_j(k) = \langle \tilde{u}_{jk}^R | \tilde{u}_{jk}^R \rangle^{-1}$, where $|\tilde{u}_{jk}^R\rangle$ are the normalized right eigenstates [38]. Phase rigidity measures the amount of mixing of two states near an EP. It vanishes at the EP according to a power-law behavior. For a normal EP, it is well known that the exponent of the power of the phase rigidity is 1/2. To verify this, we plot the phase rigidity of the second, third, and fourth bands for the case $\Gamma = 1.0$ in Fig. 5(a). It is found that phase rigidities for the states on the second and on the third bands merge and vanish at the *M* spectral singularity, whereas the phase rigidity for the states on the third and on fourth bands merge and vanish at the *S* spectral singularity. In the inset of Fig. 5(b), we show the log-log plot of the phase rigidity function near the *S* spectral singularity. A slope of 1/2 is clearly shown, indicating the singular behavior of a normal EP. However, the phase rigidities for the second, third and fourth bands near the *R* singularity exhibit a very different behavior, as shown in Fig. 5(b). We see that the phase rigidities for the states on all these three bands merge and vanish at the *R* point, indicating that there exist two defective states at this point. The log-log plot of the phase rigidity near this point gives a linear line with slope 2/3 as shown in the inset of Fig. 5(b). This is consistent with the prediction of Refs. [26,27].

**Section IV. Band Inversion in 1D *PT* symmetric PC**

Band inversion in Hermitian systems such as the Su-Schrieffer-Heeger model [39], topological insulators [40], and photonic crystals [41] has played a central role in band topological properties. Band inversion is known to change the band topological invariants such as Chern number in 2D and Zak phase in 1D and give rise to surface or interface states [39-42]. Here we are interested to know whether this concept can be generalized to the case of non-Hermitian systems. To investigate this, we focus on the third gap in Fig. 1(b). A band inversion can be induced by varying the filling ratio $d_a/\Lambda$. In the absence of gain and loss, the third gap closes when a critical filling ratio $(d_a/\Lambda)_c = 0.5064$ is reached as shown in Fig. 6(b). From the calculation of Zak phase for the band below the gap, we find that $\theta_z = \pi$ when $(d_a/\Lambda) < (d_a/\Lambda)_c$ and $\theta_z = 0$ when $(d_a/\Lambda) > (d_a/\Lambda)_c$. Away from the critical filling ratios, the gap increases and can be closed by applying a non-zero $\Gamma$. Two red lines shown in Fig. 6(b) denote the lines of spectral singularities on which the gap is closed. Below these red lines, the system is in a *PT*-exact phase which means that the gap is still open. Above these lines, the gap is closed and the system is in the *PT*-broken phase. We calculate the Zak phase for the two points marked by the two white stars in Fig. 6(b). The corresponding complex band structures are shown in Figs. 6(a) and 6(c), respectively. In general, bands can be entangled and the non-Abelian Zak phase can be calculated by using the following bilinear product [38,43]

$$\theta_{naz} = \text{Im}\left[\log\det\left(\prod_k M_{mn}^{(\mathbf{k},\mathbf{b})}\right)\right], \tag{4.1}$$

where $M_{mn}^{(\mathbf{k},\mathbf{b})} = \langle \tilde{u}_{m,\mathbf{k}}^L | \tilde{u}_{n,\mathbf{k}+\mathbf{b}}^R \rangle$, $m,n = 1,\cdots,Z$ and $Z$ is the dimension of $M^{(\mathbf{k},\mathbf{b})}$ which labels the number of entangled bands. In the case of *PT*-exact phase, the bands concerned are separated and we take $Z = 1$. The calculated Zak phases for the bands in Figs. 6(a) and 6(c) are indicated on each band. It is interesting to see that all Zak phases are still either $0$ or $\pi$, as in the case of Hermitian systems. And the bands immediately below and above the third gap remain unchanged from the case of $\Gamma = 0$. Thus, we can generalize the concept of band inversion from a Hermitian system to a non-Hermitian one as long as the system is in the *PT*-exact phase. In the *PT*-broken phase when the third gap is closed, the bands below and above become entangled. The total non-Abelian Zak phase calculated by using Eq. (4.1) with $Z = 2$ is no longer a topological invariant. Its value changes continuously with $\Gamma$. Thus, a topological transition (orange arrow line in Fig. 6(b)) can still be defined when the system goes from blue region to dark-gray region even though it has traversed a *PT*-broken phase.

**Section V. Conclusions**

In conclusion, we have found two types of *PT* phase diagrams in 1D *PT*-symmetric PCs. In one type, the *PT*-broken phase becomes an island in the $(k,\Gamma)$ phase space encircled by a ring of spectral singularities, surrounded by a continuous domain of *PT*-exact phase. In this type of *PT*-symmetry transition, Bloch eigenstates with a fixed $k$ will first enter from a *PT*-exact phase to a *PT*-broken phase as the loss/gain parameter is increased but will re-enter into a *PT*-exact phase as the loss/gain is further increased. Another type is the coalescence of spectral singularities. In this type of *PT*-symmetry transition, increase of loss/gain will induce spectral singularities at the zone center and zone boundary, and the spectral singularities move towards each other and merge to form a line of spectral singularity in the $(k,\Gamma)$ phase space and the spectral singularity is of a higher order singularity than those found in ordinary exceptional points. Finally, it is shown that the concept of topological transition as a band gap closes and re-opens due to band inversion can be generalized to non-Hermitian systems as long as the system is still in the *PT*-exact phase.

**Acknowledgements**

This work is supported by Hong Kong Research Grants Council (grant no. AoE/P-02/12).

**References**

[1] Carl M. Bender, Dorje C. Brody, and Hugh F. Jones, Phys. Rev. Lett. **89**, 270401 (2002).

[2] Nimrod Moiseyev, *Non-Hermitian Quantum Mechanics* (Cambridge University Press, Cambridge, 2011).

[3] B. Peng, S. K. Özdemir, S. Rotter, H. Yilmaz, M. Liertzer, F. Monifi, C. M. Bender, F. Nori, and L. Yang, Science **346**, 328 (2014).

[4] K. G. Makris, R. El-Ganainy, D. N. Christodoulides, and Z. H. Musslimani, Phys. Rev. Lett. **100**, 103904 (2008).

[5] Christian E. Rüter, Konstantinos G. Makris, Ramy El-Ganainy, Demetrios N. Christodoulides, Mordechai Segev, and Detlef Kip, Nature Physics **6**, 192 (2010).

[6] Alois Regensburger, Christoph Bersch, Mohammad-Ali Miri, Georgy Onishchukov, Demetrios N. Christodoulides, and Ulf Peschel, Nature **488**, 167 (2012).

[7] Wenjie Wan, Yidong Chong, Li Ge, Heeso Noh, A. Douglas Stone, and Hui Cao, Science **331**, 889 (2011).

[8] M. Brandstetter, M. Liertzer, C. Deutsch, P. Klang, J. Schöber, H. E. Türeci, G. Strasser, K. Unterrainer, and S. Rotter, Nature Communications **5**: 4034 (2014).

[9] Liang Feng, Zi Jing Wong, Ren-Min Ma, Yuan Wang, and Xiang Zhang, Science **346**, 972 (2014); Hossein Hodaei, Mohammad-Ali Miri, Matthias Heinrich, Demetrios N. Christodoulides, and Mercedeh Khajavikhan, Science **346**, 975 (2014).

[10] Li Ge, Y. D. Chong, and A. D. Stone, Phys. Rev. A **85**, 023802 (2012).

[11] Y. D. Chong, Li Ge, and A. D. Stone, Phys. Rev. Lett. **106**, 093902 (2011).

[12] Xue-Feng Zhu, Yu-Gui Peng, and De-Gang Zhao, Optics Express **22**, 18401 (2014).

[13] Romain Fleury, Dimitrios L. Sounas, and Andrea Alu, Phys. Rev. Lett. **113**, 023903 (2014).

[14] Jan Wiersig, Phys. Rev. Lett. **112**, 203901 (2014).

[15] Bo Peng, Sahin Kaya Özdemir, Fuchuan Lei, Faraz Monifi, Mariagiovanna Gianfreda, Gui Lu Long, Shanhui Fan, Franco Nori, Carl M. Bender, and Lan Yang, Nature Physics **10**, 394 (2014).

[16] Changming Huang, Fangwei Ye, Yaroslav V. Kartashov, Boris A. Malomed, and Xianfeng Chen, Optics Letters **39**, 5443 (2014).

[17] Philipp Ambich, Konstantinos G. Makris, Li Ge, Yidong Chong, A. Douglas Stone, and Stefan Rotter, Phys. Rev. X **3**, 041030 (2013).

[18] Li Ge, and A. D. Stone, Phys. Rev. X **4**, 031011 (2014).

[19] Nadav Gutman, Andrey A. Sukhorukov, Y. D. Chong, and C. Martijn de Sterke, Optics Letters **38**, 4970 (2013).

[20] Xuefeng Zhu, Hamidreza Ramezani, Chengzhi Shi, Jie Zhu, and Xiang Zhang, Phys. Rev. X **4**, 031042 (2014).

[21] Liang Feng, Ye-Long Xu, William S. Fegadolli, Ming-Hui Lu, José E. B. Oliveira, Vilson R. Almeida, Yan-Feng Chen, and Axel Scherer, Nature Materials **12**, 108 (2012).

[22] Ming Kang, Fu Liu, and Jensen Li, Phys. Rev. A **87**, 053824 (2013).

[23] Yong Sun, Wei Tan, Hong-qiang Li, Jensen Li, and Hong Chen, Phys. Rev. Lett. **112**, 143903 (2014).

[24] Mark Lawrence, Ningning Xu, Xueqian Zhang, Longqing Cong, Jiaguang Han, Weili Zhang, and Shuang Zhang, Phys. Rev. Lett. **113**, 093901 (2014).

[25] C. Dembowski, H.-D. Gräf, H. L. Harney, A. Heine, W. D. Heiss, H. Rehfeld, and A. Richter, Phys. Rev. Lett. **86**, 787 (2001); C. Dembowski, B. Dietz, H.-D. Gräf, H. L. Harney, A. Heine, W. D. Heiss, and A. Richter, Phys. Rev. Lett. **90**, 034101 (2003).

[26] Soo-Young Lee, Jung-Wan Ryu, Sang Wook Kim, and Yunchul Chung, Phys. Rev. A **85**, 064103 (2012).

[27] Kun Ding, Guancong Ma, Meng Xiao, Z. Q. Zhang, and C. T. Chan, arXiv:1509.06886 (2015).

[28] Hadiseh Alaeian, and Jennifer A. Dionne, Phys. Rev. B **89**, 075136 (2014).

[29] Alexander Szameit, Mikael C. Rechtsman, Omri Bahat-Treidel, and Mordechai Segev, Phys. Rev. A **84**, 021806 (2011).

[30] Hamidreza Ramezani, Tsampikos Kottos, Vassilios Kovanis, and D. N. Christodoulides, Phys. Rev. A **85**, 013818 (2012).

[31] Vassilios Yannopapas, Phys. Rev. A **89**, 013808 (2014).

[32] Francesco Monticone, and Andrea Alu, Phys. Rev. Lett. **112**, 213903 (2014).

[33] Bo Zhen, Chia Wei Hsu, Yuichi Igarashi, Ling Lu, Ido Kaminer, Adi Pick, Song-Liang Chua, John D. Joannopoulos, and Marin Soljačić, Nature **525**, 354 (2015).

[34] K. Sakoda, *Optical Properties of Photonic Crystals* (Springer, Berlin, 2005).

[35] Yariv Amnon, and Yeh Pochi, *Optical Waves in Crystals: Propagation and Control of Laser Radiation* (John Wiley & Sons, 1984).

[36] COMSOL Multi-physics 3.5, developed by COMSOL Inc. (Burlington, USA, 2008).

[37] Alex Figotin, and Ilya Vitebskiy, Chapter 3, *Magnetophotonics: From Theory to Applications* (Springer, 2013).

[38] I. Rotter, J. Phys. A: Math. Theor. **42**, 153001 (2009).

[39] W. P. Su, J. R. Schrieffer, and A. J. Heeger, Phys. Rev. Lett. **42**, 1698 (1979).

[40] C. L. Kane, and E. J. Mele, Phys. Rev. Lett. **95**, 146802 (2005).

[41] Meng Xiao, Z. Q. Zhang, and C. T. Chan, Phys. Rev. X **4**, 021017 (2014).

[41] Ling Lu, John D. Joannopoulos, and Marin Soljačić, Nature Photonics **8**, 821-829 (2014).

[43] A. A. Soluyanov, and D. Vanderbilt, Phys. Rev. B, **85**, 115415 (2012).

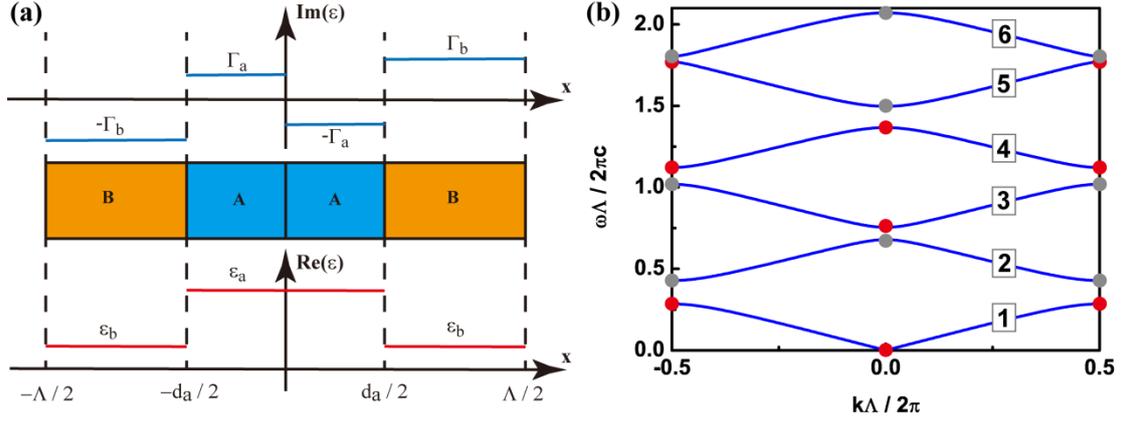

Figure 1. (color online) (a) Schematic picture of the unit cell of a *PT*-symmetric photonic crystal, and the profile of real/imaginary parts of the dielectric constants ($\text{Re}(\varepsilon)/\text{Im}(\varepsilon)$) in one unit cell. (b) Calculated band structure for the photonic crystal without loss or gain ($\text{Im}(\varepsilon)=0$, solid blue line) with system parameters $d_a = 0.42\Lambda$, $d_b = 0.58\Lambda$, $\varepsilon_a = 3.8$, $\varepsilon_b = 1$ and $\mu_a = \mu_b = 1$. The solid red circle marks an even mode ($u_{nk}(x) = u_{nk}(-x)$), and the solid grey marks an odd mode ($u_{nk}(x) = -u_{nk}(-x)$). Band numbers are indicated on each band.

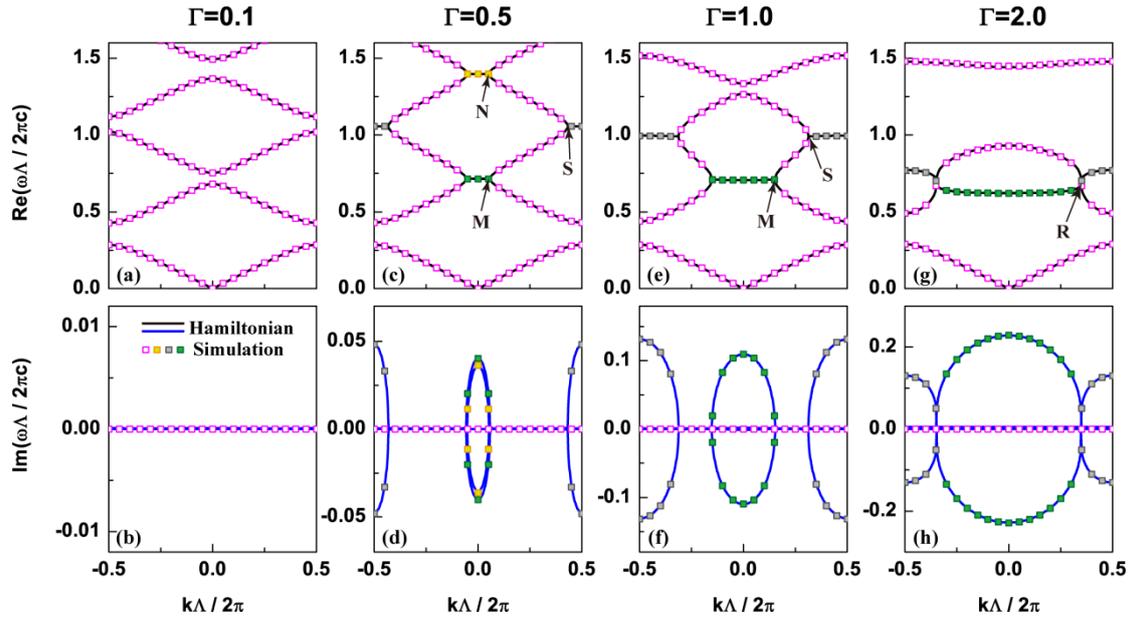

Figure 2. (color online) Complex band structures at $\Gamma(=\Gamma_a=\Gamma_b)=0.1$ (a,b), $\Gamma(=\Gamma_a=\Gamma_b)=0.5$ (c,d), $\Gamma(=\Gamma_a=\Gamma_b)=1.0$ (e,f), and $\Gamma(=\Gamma_a=\Gamma_b)=2.0$ (g,h). All the other parameters are the same as shown in Fig. 1(b). The top four panels (a,c,e,g) plot the real parts of the eigenfrequencies, and the bottom four panels (b,d,f,h) plot the imaginary parts of the eigenfrequencies. Solid lines are calculated by Hamiltonian (2.5) and all the dots are calculated by full wave simulations, in which magenta dots label *PT*-exact states, and green/gray/yellow dots label *PT*-broken states. The letters *M, S, N,* and *R* denote the spectral singularity points.

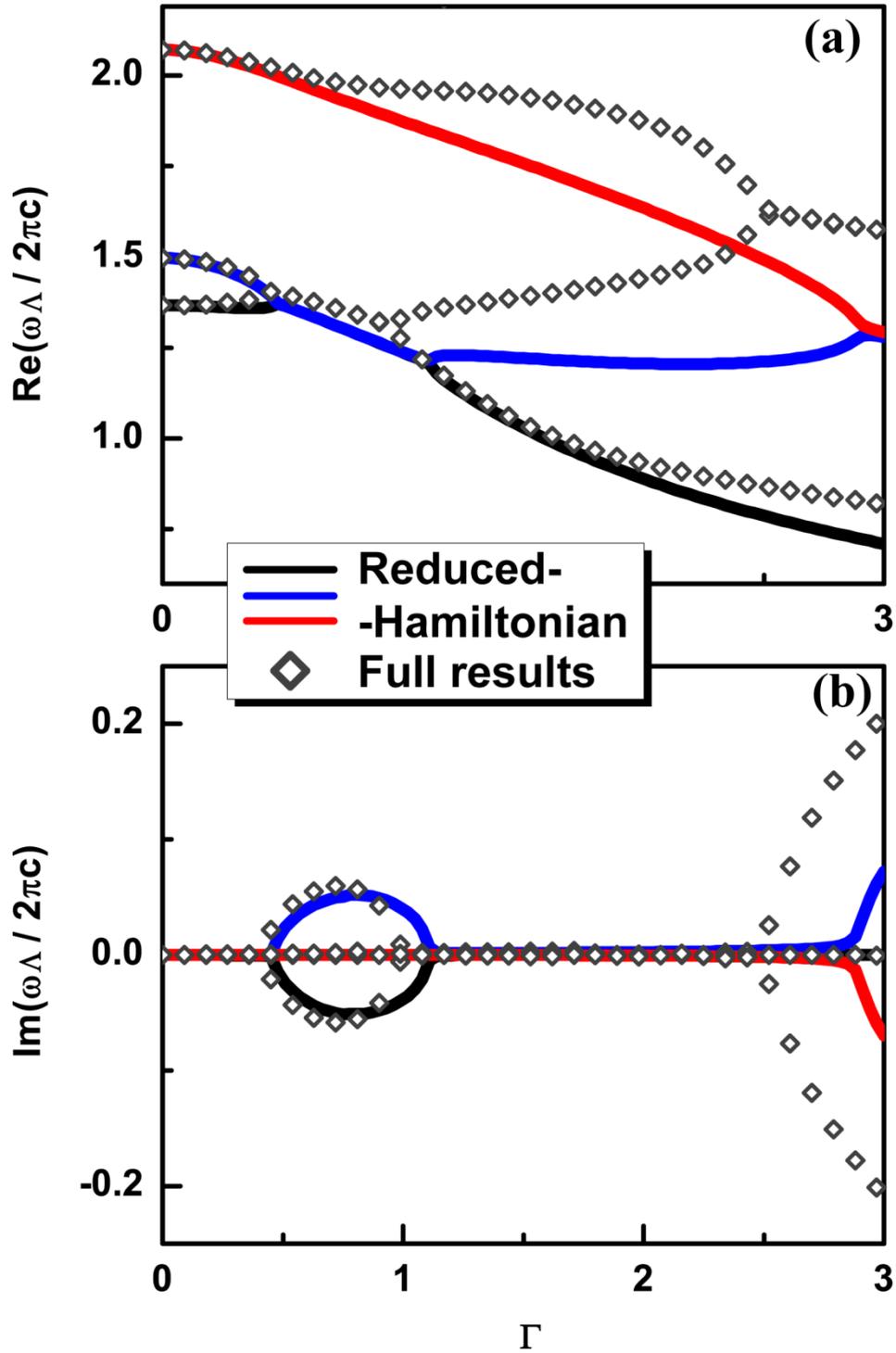

Figure 3. (color online) (a) Real parts, and (b) imaginary parts of eigenfrequencies for the states at $k\Lambda/2\pi = 0$ on the fourth, fifth, and sixth bands as functions of $\Gamma$, in which diamond dots are calculated by the full Hamiltonian and solid lines are obtained by using the $3\times 3$ reduced Hamiltonian.

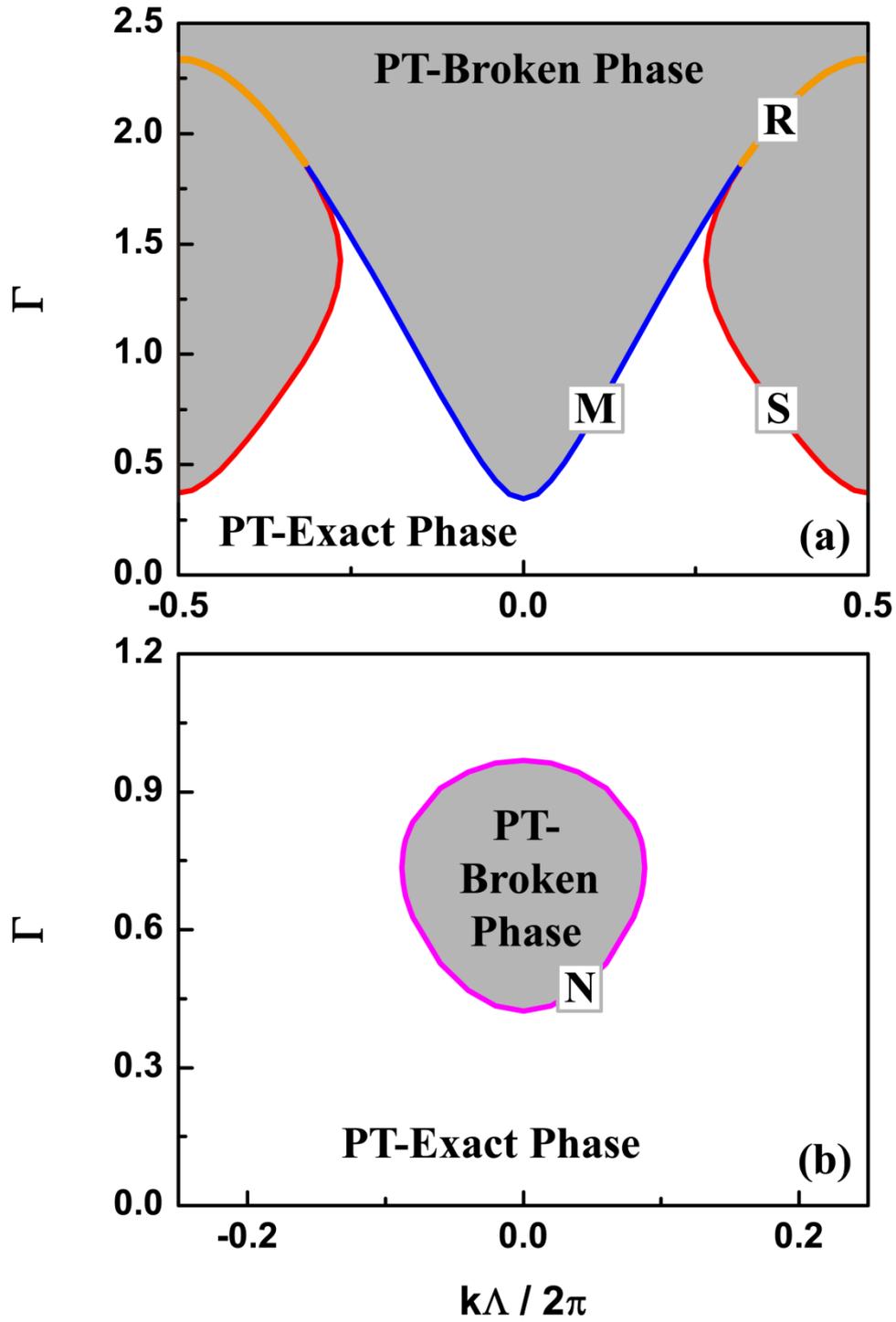

Figure 4. (color online) Trajectories of the spectral singularities in $(k, \Gamma)$ space for the parameters given in Fig. 1(b): (a) $M$ (blue line), $S$ (red line), $R$ (yellow line) and (b) $N$ (magenta line). Gray region stands for the *PT*-broken phases, and white region stands for the *PT*-exact phases.

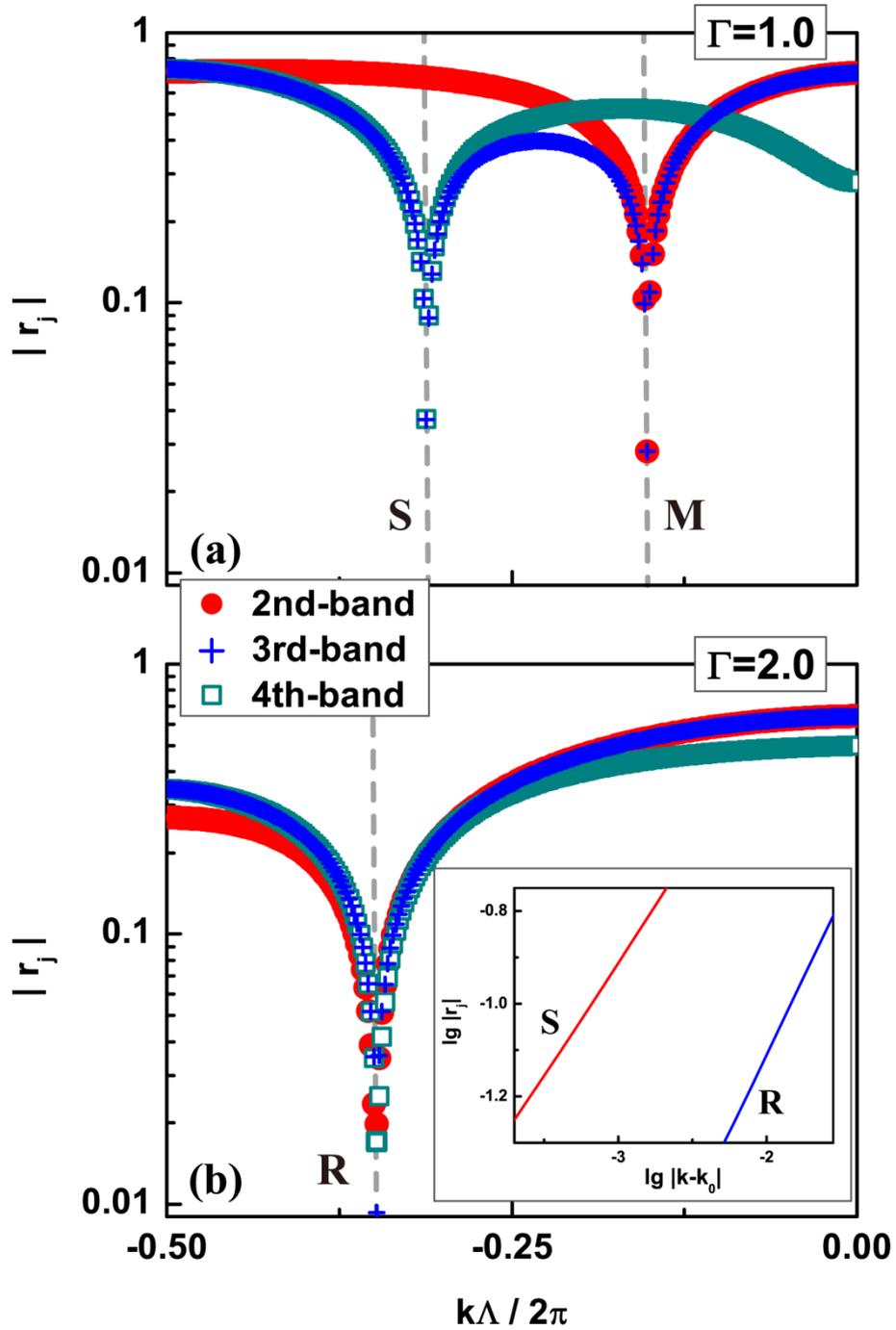

Figure 5. (color online) Phase rigidities of the states on the second-band (red dots), third-band (blue dots) and fourth-band (dark-cyan dots) as functions of Bloch $k$ for (a) $\Gamma(=\Gamma_a=\Gamma_b)=1.0$, and (b) $\Gamma(=\Gamma_a=\Gamma_b)=2.0$. The inset shows a log-log plot of phase rigidity versus $|k-k_0|$ for the spectral singularities $S$ (red line) and $R$ (blue line) respectively.

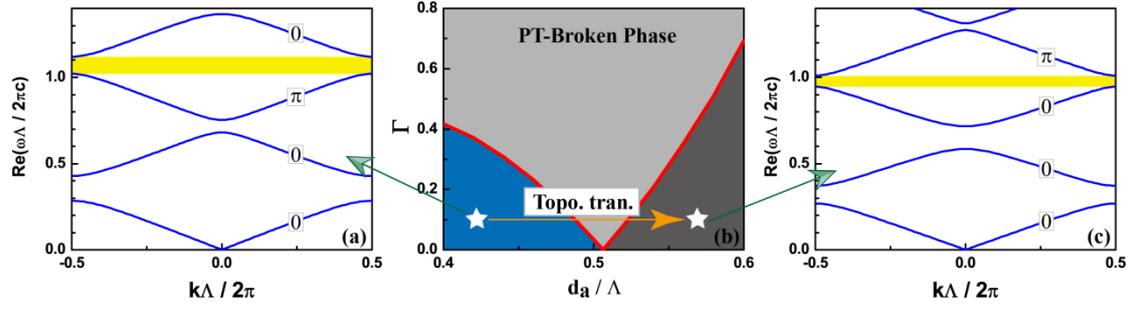

Figure 6. (color online) The central panel (b) shows the trajectory of the spectral singularity $S$ in $(d_a/\Lambda, \Gamma)$ space (red lines) for parameters $\varepsilon_a = 3.8$, $\varepsilon_b = 1$ and $\mu_a = \mu_b = 1$. Light gray region stands for the *PT*-broken phase, and blue/dark-gray regions are the *PT*-exact phases. Typical complex band structures before and after the band inversion are shown in (a) and (c) for two points shown by white stars. The calculated Zak phase for each band is also shown. The yellow regions in (a) and (c) highlight the concerned gap. The orange arrow line in (b) shows the topological transition from blue region to dark-gray region.